# High-speed and wide-field nanoscale table-top ptychographic EUV imaging and beam characterization with a sCMOS detector

**WILHELM ESCHEN,**[1,2,3*] **CHANG LIU,**[1,2,3] **DANIEL S. PENAGOS MOLINA,**[1,2,3] **ROBERT KLAS,**[1,2,3,4] **JENS LIMPERT**[1,2,3,4] **AND JAN ROTHHARDT**[1,2,3,4]

[1] *Helmholtz-Institute Jena, Fröbelstieg 3, 07743 Jena, Germany*
[2] *GSI Helmholtzzentrum für Schwerionenforschung, Planckstraße 1, 64291 Darmstadt, Germany*
[3] *Institute of Applied Physics, Abbe Center of Photonics, Friedrich-Schiller-University Jena, Albert Einstein-Straße 15, 07745 Jena, Germany*
[4] *Fraunhofer Institute for Applied Optics and Precision Engineering, Albert-Einstein-Str. 7, 07745 Jena, Germany*
*\*wilhelm.eschen@uni-jena.de*

**Abstract:** We present high-speed and wide-field EUV ptychography at 13.5 nm wavelength using a table-top high-order harmonic source. By employing a scientific complementary metal oxide semiconductor (sCMOS) detector the scan time for sub-20 nm high-resolution measurements were significantly reduced by up to a factor of five. The fast frame rate of sCMOS enables wide-field imaging with a field of view of 100 µm x 100 µm with an imaging speed of 4.6 Mpix/h. Furthermore, fast EUV wavefront characterization is employed using a combination of the sCMOS detector with orthogonal probe relaxation.

## 1. Introduction

X-ray microscopy has seen tremendous progress in recent years fueled by the development of modern lensless imaging methods [1,2]. In particular, ptychography [3–5], which is a scanning version of coherent diffractive imaging, has emerged as a powerful method at synchrotron facilities [6]. In ptychography the object is scanned through an illuminating beam while keeping an overlap between neighboring illuminated regions. The resulting diversity of the recorded diffraction patterns behind the object allows for numerical reconstruction of the amplitude and phase of the object's transmission function and of the illumination, which is usually referred to as probe. Ptychography offers significant advantages compared to other imaging techniques. First, ptychography allows surpassing the resolution limits imposed by optical aberrations, due to its lensless nature. Second, the large amount of diverse data results in stable convergence of the reconstruction algorithm. Finally, ptychography reconstructions give access to quantitative amplitude and phase contrast due to the deconvolution of the probe and the object's angular spectrum.

In addition, ultrafast laser technology has undergone tremendous progress in recent years, which facilitated the development of table-top high photon flux coherent EUV sources based on high-harmonic generation (HHG). It was shown that HHG sources can reach well into the X-ray region [7] and can provide a photon flux comparable to Synchrotron facilities [8] in the extreme ultraviolet (EUV). Ultimately, the combination of HHG-based XUV sources with ptychography will facilitate nanoscale phase-sensitive imaging on a laboratory scale.

The EUV range offers a unique combination of resolution, element specificity, and penetration depth. For this reason, coherent diffractive imaging driven by compact HHG sources has gained increasing attention [9]. The use of such HHG sources for EUV ptychography [10], in spite of being in its infancy, has led to high-resolution imaging with sub-20 nm resolution [11,12]. The excellent element specificity of EUV radiation has been used for material-specific imaging in reflection [13,14] and transmission [11]. This opens up a wide range of potential applications, which include actinic metrology for the semiconductor industry at 13.5 nm [15–19], high-resolution imaging of biological specimens [20,21], and the investigation of specific scientific issues [22]. In addition, EUV ptychography is used for high-resolution and broadband wavefront sensing [23,24], which is of particular interest for the characterization of broadband HHG sources [25] and EUV optics [26].

However, until now, the performance of EUV ptychography has been limited by the power of the coherent EUV sources, detector noise, and especially the slow readout rate of the charge-coupled device (CCD) detectors used. Although short exposure times in the sub-second range are often used, the complete measurement time, which includes read-out of the CCD, data transfer and movement of the positioners, is usually much longer. Often only the accumulated exposure time is reported, but the read-out time is not. Since EUV ptychography experiments are performed in different experimental geometries and at a broad range of wavelengths, the achieved spatial resolution and measured field of view vary greatly. Nevertheless, the imaging speed can be compared if calculated in megapixels per hour (Mpix/h), which is independent of the experimental geometry and wavelength [27]. It should be noted that the imaging speed is of course also dependent on the available photon flux, which for HHG sources significantly drops for shorter wavelengths [28]. To the best of our knowledge at the technologically relevant wavelength of 13.5 nm, the imaging speed was so far below 1 Mpix/h [11,12]. Today, sCMOS (scientific complementary metal oxide semiconductor)-type detectors for the EUV and soft X-ray spectral range have become commercially available and have shown excellent quantum efficiency and noise performance [29]. In the soft X-ray range, the benefits of sCMOS sensors have already been demonstrated at synchrotron beamlines [30,31].

In this paper, we employ a sCMOS detector in a table-top EUV ptychography setup for the first time and demonstrate high-speed, high-resolution, and wide-field EUV ptychography at significantly increased speed. The novel detector brings many benefits compared to conventional CCDs. The improved noise performance allows a reduction of the exposure time without sacrificing the signal-to-noise ratio and spatial resolution. Further, the high frame rate (> 20 frames per second for the settings used here) enables fast scanning of large areas, which was previously not feasible due to the long read-out time of the used CCDs. For sub-20 nm-resolution measurements, a reduction of the total measurement time by more than a factor of 2 compared to previous measurements [11] is demonstrated without sacrificing resolution. Further, we demonstrate high-speed wide-field imaging of a 100 µm × 100 µm large area with a half-pitch resolution of 79 nm in under 21 minutes, which corresponds to 4.6 Mpix/h - 5 times faster than previously reported with the same EUV source and imaging setup [11]. Finally, the high frame rate of the detector is ideal for fast and high-resolution characterization of EUV wavefronts. We demonstrate wavefront characterization in a single scan in under 80 seconds and fast, dynamic wavefront sensing by combining orthogonal probe relaxation [32] with the novel detector, which enables tracking wavefront changes on the Hertz scale.

## 2. Experimental setup

### 2.1 Setup

The ptychographic EUV microscope used here has been previously used in a similar configuration and has demonstrated high-resolution imaging of samples relevant to material sciences and biology [11,33]. In short, a ytterbium-doped fiber-based chirped-pulse amplifier emitting at 1030 nm is used. The amplifier provides 250 fs pulses with 1 mJ pulse energy at a repetition rate of 75 kHz resulting in an average power of 75 W. The ultra-short pulses are further compressed by two noble-gas-filled hollow core fibers to sub 7 fs with a remaining pulse energy of 0.4 mJ. The IR laser is subsequently focused in an argon gas jet (see Figure 1), where a broadband EUV continuum up to 100 eV is generated with a photon flux of $7 \times 10^9$ phot/s/eV at 13.5 nm (92 eV) [34]. The broadband EUV beam is separated from the high-power IR laser by four grazing incidence plates, which reflect most of the EUV radiation but transmit the driving IR beam [35]. The remaining IR light (<50 mW) is blocked by two 200 nm thick Zirconium foils. The HHG source is imaged on the sample using three multilayer mirrors (M1, M2, and M3) in a Schiefspiegler configuration [11]. The multilayer mirrors have a reflectivity of > 60% at 13.5 nm each, yielding a combined peak reflectivity of 23% and a relative bandwidth (FWHM) of $\lambda/\Delta\lambda = 60$ at 13.5 nm. The EUV mirror setup employs multilayer mirrors with different angle of incidence (AOI) on each mirror to compensate for astigmatism. Previously, for all EUV mirrors the same multilayer design was used, which results in reduced total reflectivity since the peak reflectivity of the multilayer mirrors shifts with a change of AOI. Here, a new multilayer mirror was employed with a matching peak reflectivity for the different AOI, which results in 50% more photon flux on the sample. A binary mask (Figure 1), which is used to structure the illumination, can be moved up to ~100 μm upstream of the sample. It has been shown, that a structured illumination is beneficial for ptychography [36,37]. Here we use a spiral-shaped mask, which has been previously used and improves the image quality compared to a simple pinhole [11]. The diffraction patterns are collected using a sCMOS detector (GSENSE400BSI, Andor Marana-X 11) with 2048 × 2048 pixels and a pixel size of 11 μm placed 28 mm behind the sample, supporting a diffraction-limited resolution of 18 nm. The object is reconstructed from the measured diffraction patterns using the mPIE [38] implementation of the ptyLab framework [39].

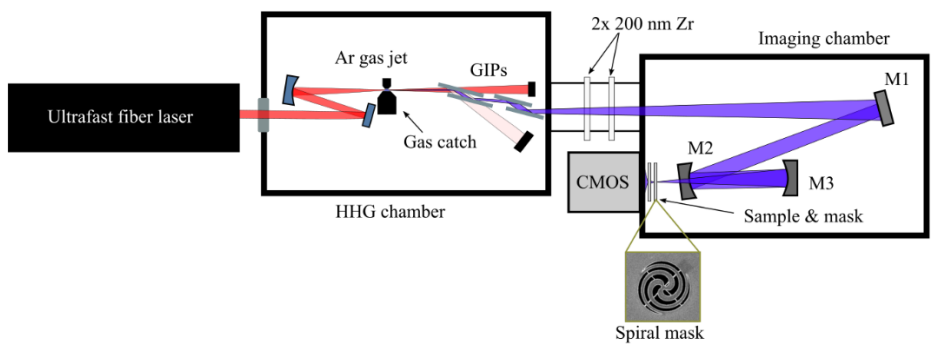

Fig. 1. High-harmonic generations (HHG) setup and ptychographic vacuum imaging chamber. An ultrafast few-cycle fiber laser with a central wavelength of 1030 nm is focused in an argon gas jet, where a broadband EUV continuum is generated. The driving infrared (IR beam is separated from the EUV radiation by four grazing incidence plates (GIPs) that transmit the IR radiation but reflect the EUV radiation. The remaining IR radiation is blocked by two 200 nm Zr foils. The HHG source is imaged on a beam structuring mask using three EUV multilayer mirrors which act also as a monochromator selecting a wavelength of 13.5 nm. The diffraction patterns are collected using a sCMOS detector.

*2.2 Benefits of sCMOS detectors for EUV ptychography*

So far CCDs were used for EUV ptychography due to their widespread availability. However, in a CCD the charges, generated by the absorbed EUV photons, are moved across the chip and are read at one corner where the electrical signal is converted by an analog-to-digital converter to counts. Thus, the speed of EUV ptychography measurements was so far limited by the readout speed of the involved CCDs. In a CMOS sensor, on the other hand, there are several transistors behind each pixel. Thus, each pixel can be read out separately, resulting in considerably higher frame rates compared to CCDs. Here we compare a state-of-the-art CCD (Andor iKON-L) with a novel sCMOS detector (Andor Marana-X 11) for use in EUV ptychography at 13.5 nm. The sCMOS detector consists of the backside-illuminated GSENSE400BSI sensor (2048×2048 pixel, 11 μm × 11 μm pixel size), which shows a high quantum efficiency (>90%) in the EUV / soft X-ray region and a low read-out noise of 1.6 electrons [29,40].

First, the noise performance of the CCD and sCMOS detectors is compared. A single ptychography measurement usually contains more than 100, and can even exceed thousands of measured diffraction patterns. Therefore, EUV CCDs are typically operated in a trade-off between readout speed and readout noise. For previous measurements, a readout rate of 1 MHz was chosen [11,41]. Using a 2×2 on-chip binning, the resulting frame rate for the full chip is 0.67 Hz with a read-out noise of 9.7 electrons (RMS). In comparison, the measured readout noise of the 16-bit mode of the sCMOS detector is only 1.6 electrons (RMS) (see Figure 2 b). However, the thermal noise of the sCMOS detector is significantly larger compared to a standard CCD, which results in thermal noise dominating over readout noise for long exposure times. For thermal noise measurements, the sCMOS detector was cooled to -45°C, and dark images were acquired for exposure times ranging from 1 ms to 60 s using the 16-bit mode of the detector. The noise level was characterized by calculating the standard deviation in a central area with the size of 200 × 200 pixels (white box indicated in Figure 2 a). Please note that the detector noise increases at the very edge of the detector. This region was not included in the noise analyses. The corresponding noise values are shown in Figure 2 b for exposure times ranging from 1 ms to 60 s for 1×1, 2×2, and 4×4 binning. For comparison, the read-out noise (9.7 electrons (RMS)) of the state-of-the-art CCD operated under typical conditions is shown in Figure 2 b as well. Since the CCD exhibits substantially less thermal noise (0.0002 electrons/s/pxl at -100°C) it can be neglected. We find that the sCMOS detector surpasses the CCD even for long exposure times using 1x1 binning, while still offering a faster read-out.

It should be noted that with sCMOS detectors binning increases the noise, which is due to the separated read-out channels, while for CCDs on-chip binning does not increase the noise level. Especially for high binning values (e.g. 4x4 binning) the noise is increased significantly for the sCMOS detector and might even be worse compared to the situation using the CCD (cf. Figure 4 b). For instance, using 2×2 binning and an exposure time of 60 s results in better noise performance of the CCD compared to the sCMOS detector. Whereas for 4x4 binning already an exposure time of 10 s shows an improved noise performance of the CCD. However, EUV ptychography usually operates in the far field, which means that the required sampling of the diffraction pattern can usually be achieved by choosing an adequate sample to detector distance. Therefore, instead of binning, the detector can also be moved closer to the sample which results in the same Fourier-space sampling of the diffraction pattern. Hence, we conclude that the sCMOS detector shows an improved noise performance for most practical cases, except for large binning values (e.g. 4x4) which in principle can be avoided by adjusting the experimental geometry.

Furthermore, to be able to measure photons up to the highest diffraction angles, high-dynamic range imaging of the diffraction patterns is often used. For this purpose, multiple images with increasing exposure time are stitched to a single high-dynamic range diffraction pattern [42]. For the longer exposure times, the diffraction patterns are usually overexposed in the center where the missing information is recovered by the short exposure measurement. In the overexposed area this can lead to blooming artifacts, where charges on the CCD are not properly read out anymore. This is especially problematic for weakly scattering samples. In this case, the diffraction patterns can exhibit long streaks in the vertical direction that corrupt further areas. To demonstrate this effect, an overexposed diffraction pattern of the beam structuring element with $\approx 6 \cdot 10^6$ measured photons is shown in Figure 2 c. The long streaks, which corrupt significant areas of the diffraction data, are visible in the center of Figure 2 d. The streaks are problematic for ptychography because they can cover areas of the diffraction patterns with low intensity which are not measured with sufficient signal-to-noise ratio by the short exposure measurements. Therefore, it is often not possible to recover these areas at all or with a sufficient signal-to-noise ratio.

For comparison, a diffraction pattern of the same structure is measured with the sCMOS detector with a similar number of detected photons (Figure 2 d). In this image, the streaks do not appear since the charges are not moved across the sCMOS detector but are directly converted at each pixel. This makes it much easier to acquire high-quality ptychographic HDR images. A second mitigation strategy is to block the overexposed area with a beam stop [41]. However, this usually requires mechanical movement of an absorbing structure which increases the complexity of the ptychography setup and, at the same time, increases the total measurement time.

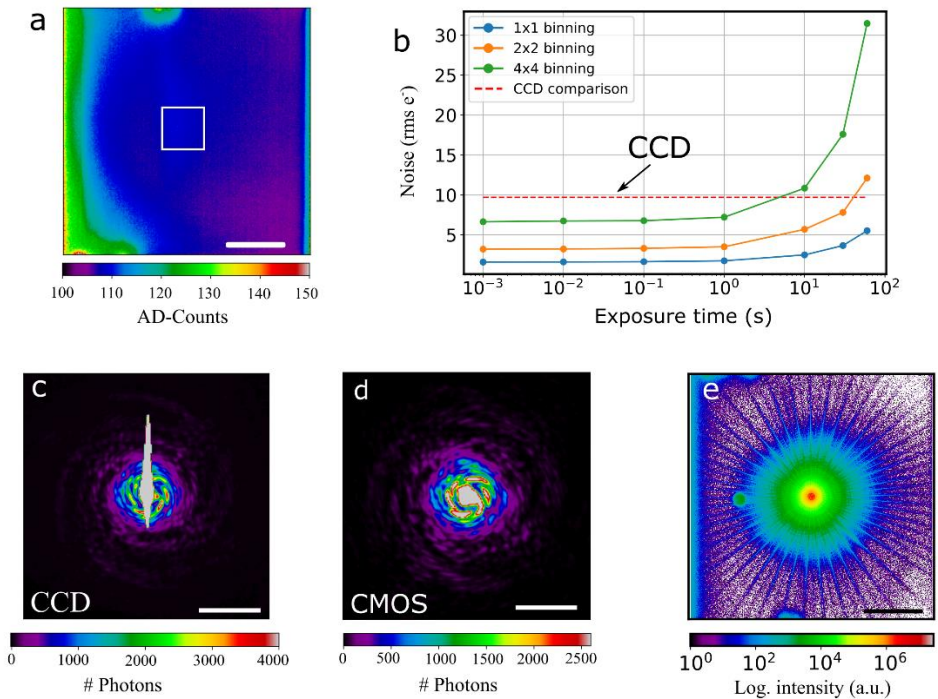

Fig. 2. a) Measured dark image of the CMOS sensor (Marana-X 11) using an exposure time of 20 s. In b) the rms-noise-level of dark images for varying exposure times ranging from 1 ms to 60 s and for different binning is plotted. The read-out noise of the previously used CCD (Andor iKon-L) is indicated by a red dashed line. Since the thermal noise of the CCD is low, it can be neglected for the exposure times shown here. c) shows an overexposed diffraction pattern containing ~$6 \cdot 10^6$ photons. In the center blooming of the CCD detector is visible which corrupts mainly vertical areas. In d) a diffraction pattern of the same structure is depicted, which has been measured with a sCMOS detector (Andor Marana-X 11) and shows less blooming and vertical streaks. e) shows the averaged high-dynamic range diffraction pattern of a ptychography scan using 101 positions. The scale bar in a) and e) corresponds to 500 pixels of the detector, while the scale bar in c) and d) corresponds to 50 pixels.

## 3. High-speed and wide-field EUV ptychography

The benefits of the sCMOS detector are demonstrated by performing ptychography on a Siemens star resolution test chart. The high frame rate of the sCMOS detector favors short exposure times and therefore allows the scanning of large areas with reduced resolution in a short time. The Siemens star was scanned by 1004 positions spaced by 3 μm. Only the center of the sCMOS detector (512 × 512 pixels) was read out during the measurement. For each scan position, an exposure time of 1 s was chosen and the whole measurement took 1249 seconds (~21 minutes). For every position, readout and data transfer took 0.12 seconds, while the movement of the positioners took 0.11 seconds. For the reconstruction 4 incoherent modes [43] were used for the probe, resulting in 74% power content in the most dominant mode. The resulting reconstruction of the object is depicted in Figure 3, showing a high-quality image on a field of view as large as 100 μm × 100 μm. The resolution was evaluated using the Fourier ring correlation (FRC) and the half-bit criterion (Figure 3 d) resulting in a half-pitch resolution of 79 nm which corresponds to an acquisition speed of 4.6 megapixels per hour - 5 times faster than previously reported [11].

In the next step, a second, high-resolution measurement of the center of the Siemens star was performed. For this purpose, 101 positions were acquired with an increased overlap by reducing the distance between neighboring positions to 1 μm. Two diffraction patterns with 2 s and 20 s exposure time were acquired and stitched to a single, high-dynamic range diffraction pattern for every position. Figure 2 e shows the stitched diffraction patterns averaged for all positions. Although an averaged dark image for the long exposure time (20 s) was subtracted, thermal noise at the edges of the diffraction pattern is still visible. Nevertheless, the resulting reconstruction (Figure 3 b), shows high image quality, and even the most central spokes of the Siemens star are resolved (Figure 3 c). The resolution here was estimated by FRC to be 18 nm, which is similar to previous measurements obtained with a CCD detector [11]. However, the overall measurement time was reduced from 91 min to 38 min which corresponds to a speed improvement by more than a factor of 2. Besides the lower noise level of the sCMOS detector (-65%), an increased photon flux (+50%) on the sample contributed to this improvement as well.

To quantify the obtained amplitude and phase precision of the obtained image, the amplitude and phase values of all pixels within an empty area, indicated by the red rectangle in Figure 3b, were plotted as a histogram in Figure 3e and Figure 3f. The amplitude- and phase- precision are calculated as the standard deviations, which is 2% for the amplitude and 27 mrad for the phase. Overall, the presented EUV microscope provides significantly faster imaging than previous demonstrations and allows for high-quality quantitative amplitude- and phase- imaging across multiple scales ranging from 0.1 mm to below 20 nm.

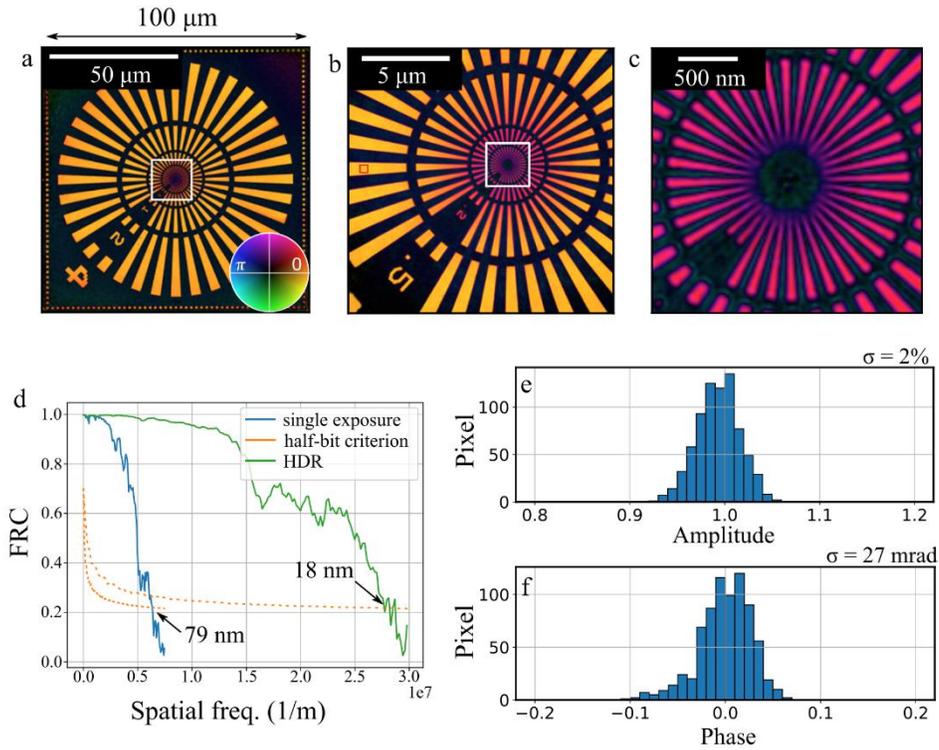

Fig. 3 a) Reconstruction of a 100 μm × 100 μm field of view of the Siemens star using 1004 measured diffraction patterns. b) High-dynamic range measurement using two exposure times for each position of the center of the Siemens star. c) Region of interest indicated in b showing the most inner spokes of the Siemens star. The brightness and hue of a, b and c encode modulus and phase. d) Fourier ring correlation (FRC) showing for the large field of view measurement (Figure 3 a) 79 nm resolution and the high-resolution measurement (Figure 3 b) 18 nm resolution. e and f show the amplitude and phase distribution in a histogram for the red rectangle indicated in Figure 3 b. The precision is given by the standard deviation which was calculated to be 2% for the amplitude and 27 mrad for the reconstructed phase.

## 4. High-speed ptychographic EUV wavefront sensing

In addition to high-resolution imaging, ptychography is commonly used to characterize coherent short-wavelength wavefronts. It benefits especially from the high spatial resolution, compared to other methods [44,45]. Furthermore, ptychography allows the reconstruction of broadband [23] or partially spatial coherent beams [46], which provides additional information. However, since ptychography usually requires multiple diffraction patterns (typically more than 100) the measurement of a single EUV wavefront took multiple minutes at HHG setups so far. In contrast, the alignment of optical systems or the optimization of the high-order harmonic parameter space often requires faster feedback. Here, the measurement of a single wavefront within 78 s is demonstrated, by characterizing the aberrated wavefront of the employed Schiefspiegler EUV telescope. For this purpose, the beam structuring mask was replaced by a pinhole with a diameter of 12 μm (Figure 4 a) to block EUV stray light.

The focal spot with a diameter of 4 μm × 3 μm was placed in the center of the aperture. 146 positions spaced by 1.5 μm were recorded with an exposure time of 300 ms each and an overhead time (read-out, data transfer, and movement of the positioners) of 230 ms per position. This results in a total measurement time of 78 s, which is significantly faster compared to previous measurements where usually more than 10 minutes were required for a wavefront measurement [23,25]. For the reconstruction of the object and the probe 4 spatially incoherent modes were used revealing 68% of the total power in the dominant mode, which is shown in Figure 4 b. The reconstructed beam is elliptical and shows a strong aberration of the wavefront, which we attribute to a non-perfect alignment of the Schiefspiegler configuration.

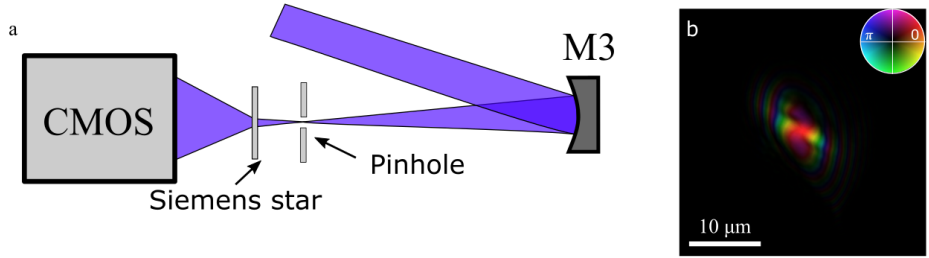

Fig. 4 a) Sketch of the wavefront measurement. Instead of a beam structuring mask, a pinhole with a diameter of 12 μm was placed in the focal plane of the EUV beam. The reconstructed complex beam is shown in b. The brightness and hue of b encode modulus and phase.

For fast alignment procedures or ultrafast studies, even faster frame rates are desired. Here we propose to use orthogonal probe relaxation ptychography (OPR) [32] for dynamical imaging of the wavefront. Usually, ptychography requires that the probe does not change during the measurement. Hence, a changing illumination reduces the image quality or can even lead to the situation that the reconstruction algorithm does not converge at all. However, the OPR method relaxes this requirement, by assuming that the illumination is not changing completely randomly but rather only slightly, which is the case for most applications. For each scan position, a distinct probe is reconstructed, but the OPR method requires that each probe can be sufficiently approximated by a coherent superposition of a limited number of eigen-modes. The eigen-modes are retrieved by a truncated singular value decomposition, which is performed on a stack that contains the probes of all positions. This method has previously been used already to account for temporal drifts of illuminating beams [11,20,32], and for shot-to-shot wavefront sensing at free-electron laser facilities [47].

Here, OPR is applied to track changes in the high-harmonic generation process. For this purpose, the argon gas jet is moved along the propagation direction during the ptychography scan (Figure 5 a), which leads to a change of the generation conditions and therefore also to a significant change of the EUV beam. Usually, these changes are only observed when the sample is moved out of the beam path and the far-field intensity-only-image of the beam is measured by the detector, which does not give information on the focal spot size or aberrations of the EUV beam.

The ptychography scan is similar to the previous measurement (Figure 4 b), but the number of positions is increased to 258, and for every 65 positions the gas jet was moved along the propagation direction of the IR-laser beam by ~250 μm. Therefore, during the single scan, four different EUV beam profiles are obtained. Using ptychography with 4 incoherent modes without OPR analysis leads, as expected due to the changes of the probe during the scan, to poor image quality (Figure 5 b).

If the OPR method is used for the reconstruction and each incoherent mode is linked by a truncated singular value decomposition in a subspace of 4 (i.e. only the 4 most dominant singular values are used), the image quality can be significantly improved (Figure 5 c) and a probe for each position is successfully reconstructed. The probe power for each position is shown in Figure 5 d, which shows a change in power for the four different nozzle positions with the highest HHG efficiency being obtained at nozzle position 2. For each nozzle position, the dominant mode is plotted in Figure 5 d-g where it is apparent that the defocus increases with an increasing axial displacement of the nozzle. To confirm that the reconstructed probes are physical, the intensity far-field image is calculated from the reconstructed probe (Figure 5 d-g labeled as 'recon') for each nozzle position respectively and compared to the corresponding empty EUV beam (sample moved out of the beam) measured on the detector (shown in Figure 5 d-g labeled as 'detector). The comparison shows that the measured far-field images are in good agreement with the numerically retrieved far-field images of the reconstructed probe modes, which indicates that the probes were reconstructed properly.

Note, that by employing the OPR method the effective frame rate for the characterization of the EUV beam is thus increased by more than 2 orders of magnitude to 2 Hz. This opens up new possibilities to study fast processes and provide rapid feedback for the alignment of optics, complex optical systems, and adaptive optics. It should however be noted, that the OPR method needs currently more than 500 iterations to converge, which results in long reconstruction times (>10 minutes) which hinders direct feedback. However, the reconstruction time can in principle be drastically reduced by starting with a pre-measured initial guess of the object which can be obtained by an initial scan of the same area without a changing probe and by employing more sophisticated high-performance computing methods [48].

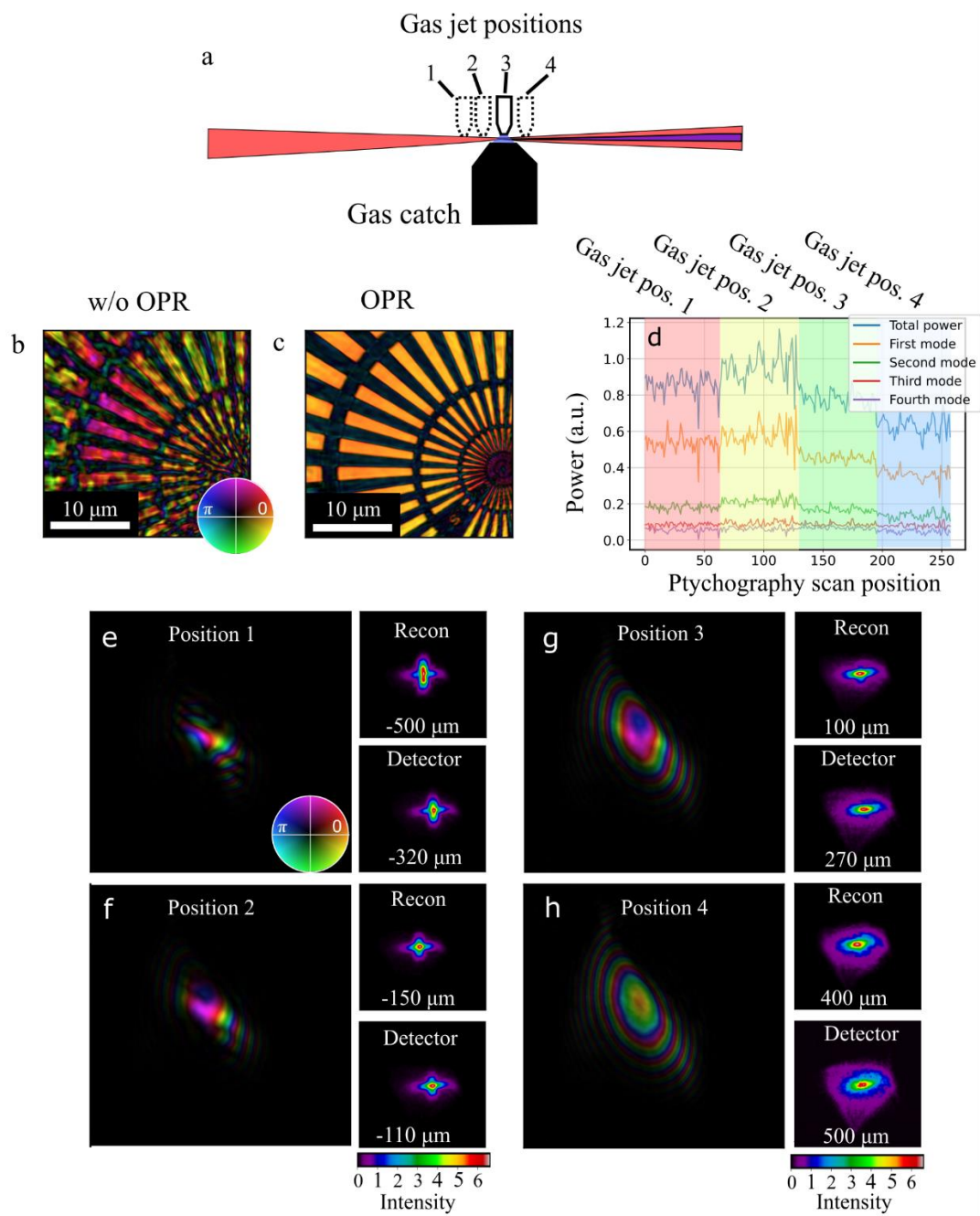

Fig. 5 a) Schematical sketch of the varying high-harmonic generation conditions. The gas jet is moved along the propagation direction during the ptychography scan. The movement of the nozzle leads to different intensities and wavefronts during the HHG process. b) Standard ptychography reconstruction using 4 incoherent modes without using orthogonal probe relaxation. c) Using 4 modes and orthogonal probe relaxation on all incoherent probe modes. d) Probe power for all positions for all incoherent probe modes. e)-f) shows the reconstructed EUV beams for four gas jet positions indicated in a). The small insets show the calculated far field pattern of the reconstructed beam labeled 'Recon' and for similar positions measured far field patterns. The brightness and hue of b, c, e, f, g, and h encode modulus and phase.

## 5. Summary and outlook

In summary, we have presented significant advances in table-top EUV imaging and beam characterization. By employing a novel commercially available sCMOS detector for ptychography combined with an improved EUV multilayer system we demonstrate quantitative EUV imaging on a 100 µm × 100 µm field of view with 4.6 Mpix/h, which is 5 times faster than previously reported with the same EUV source, while still enabling sub-20 nm resolution using high-dynamic range measurements. Further, by applying the OPR method, we demonstrated EUV beam characterization and wavefront sensing with Hz-level frame rates. While the demonstrated results at 13.5 nm wavelength are of particular interest for applications in the realm of EUV lithography, where fast scanning of large areas is of great interest, we believe that HHG-source and beam characterization at other wavelengths will benefit from our findings as well. We believe that the presented benefits of sCMOS detectors for EUV ptychography are also available for other EUV imaging modalities including 3D [49] and full-field achromatic imaging [50].

The bottleneck for high-resolution ptychographic imaging at 13.5 nm remains the limited photon flux of the available HHG sources. However, by combining sCMOS detectors with the latest mW-class HHG sources [8] and advanced multiplexing [51] or fly-scan ptychography [52], Gigapixel EUV imaging and kilohertz frame rate beam characterization appears feasible in the future. This would allow e.g. for imaging and inspection of mm²-sized areas of EUV lithography wafers and masks with nanoscale resolution.

**Funding.** The research was supported by the Thüringer Ministerium für Bildung, Wissenschaft und Kultur (2018 FGR 0080), the Helmholtz Association (Incubator Project Ptychography 4.0 and Helmholtz-Imaging Project ZT-I-PF-4-018 (AsoftXm)) and the Fraunhofer- Gesellschaft (Cluster of Excellence Advanced Photon Sources).

**Acknowledgments.** We thank Michael Steinert from the Institute of Applied Physics at the University of Jena for the preparation of the spiral-shaped mask.

**Disclosures.** The authors declare no conflict of interest.

**Data availability.** Data underlying the results presented in this paper are not publicly available at this time but may be obtained from the authors upon reasonable request.